\documentclass[10pt,a4paper]{article}
\usepackage[utf8]{inputenc}
\usepackage[english]{babel}
\usepackage{authblk}
\usepackage{amsmath}
\usepackage{amsfonts}
\usepackage{amssymb}
\usepackage{graphicx}
\usepackage{caption}
\usepackage{subcaption}
\author{Arindam Saha\thanks{Electronic address: \texttt{arindam1002@iiserkol.ac.in}}
}
\affil{\textit{Department of Physical Sciences}\\ \textit{Indian Institute of Science Education and Research Kolkata}}
\author{R. E. Amritkar\thanks{Electronic address: \texttt{amritkar@prl.res.in}}
}
\affil{\textit{Department of Theoretical Physics}\\ \textit{Physical Reasearch Laboratory Ahmedabad}}
\title{Dependence of synchronisation frequency of Kuramoto oscillators on symmetry of natural frequency distribution}
\date{}
\begin{document}
\maketitle

\begin{abstract}
Kuramoto oscillators have been proposed earlier as a model for interacting systems that exhibit synchronisation. In this article we study the difference between networks with symmetric and asymmetric distribution of natural frequencies. We first indicate that the synchronisation frequency of the oscillators is independent of the natural frequency distribution for a completely connected network. Further we analyse the case of oscillators in a directed ring-network where asymmetry in the natural frequency distribution is seen to shift the synchronisation frequency of the network. We also present an estimate of the shift in the frequencies for slightly asymmetric distributions.
\end{abstract}

\section{Introduction}

The phenomenon of synchronisation is seen in interacting oscillatory systems in nature. Most striking examples include the regular flashing of light by fireflies \cite{buck1988synchronous}, simultaneous clapping by the audience in a theatre during an applause \cite{neda2000physics,gerstner1995time}, Josephson junctions \cite{PhysRevE.57.1563,wiesenfeld1996synchronization} and chemical oscillations \cite{kuramoto2003chemical}. Collective synchronisation was first studied mathematically by Wiener \cite{wiener1966nonlinear,wiener1948cybernetics}. He realised the ubiquity of the phenomenon and speculated its involvement in generation of alpha rhythms in the brain. Unfortunately Wiener's mathematical approach based on Fourier integrals \cite{wiener1966nonlinear} has turned out to be a dead end \cite{strogatz2000kuramoto}. In 1975, Kuramoto introduced a model which took into consideration oscillators which were coupled to each other and showed the phenomenon of synchronisation for sufficient large coupling strengths.

The dynamics of a general $i^{th}$ oscillator in a system of $N$ Kuramoto oscillators is given as
\begin{equation}
\dot{\theta_i}=\omega_i + \sum_{j=1}^{N} K_{ij} \sin(\theta_{j}-\theta_{i}) ~~~ \forall i=1,2,...,N 
\label{eq: Kuramoto}
\end{equation}
where $\theta_i$ and $\omega_i$ are the phase and natural frequency of the $i^{th}$ oscillator respectively and $K_{ij}$ is the coupling strength between the $i^{th}$ and $j^{th}$ oscillators. For simplicity, we apply the mean field approximation and set $K_{ij}=\frac{K}{N}$ for all $1 \leq i,j \leq N$.

In this article we analyse the Kuramoto model with an aim to study the difference in dynamics of the oscillators as the distribution from which natural frequencies of the oscillators are chosen changes from symmetric to asymmetric form. In particular, we study the change in synchronisation frequency as the symmetry is changed under the limit of large $N$. We first analyse a complete network of oscillators and show that symmetry of natural frequency distribution has no effect on the synchronisation frequency. We then consider a network of oscillators connected in a directed ring and obtain qualitative differences between the symmetric and asymmetric cases. These differences are presented as numerical results. We also analyse the differences analytically and present an estimate of the shift in synchronisation frequency.

\section{Synchronisation in completely connected network}

To show the independence of synchronisation frequency on symmetry of natural frequency distribution, we consider the dynamical equation \eqref{eq: Kuramoto} for the completely synchronised state of oscillators under the mean field approximation.  In such a scenario, all oscillators have the same effective frequency $\dot{\theta_i}=\Omega$ and the equation gets transformed as
\begin{equation}
\Omega=\omega_i + \frac{K}{N} \sum_{j=1}^{N} \sin(\theta_{j}-\theta_{i}) ~~~ \forall i=1,2,...,N 
\label{eq: Synchronised Complete}
\end{equation}
Adding the equations for all $i$'s together gives
\begin{equation}
\Omega=\overline{\omega}_i + \frac{K}{N} \sum_{i=1}^{N}\sum_{j=1}^{N} \sin(\theta_{j}-\theta_{i})
\end{equation}
The second term in the equation drops out due to its antisymmetry in $i$ and $j$ and leads to the synchronisation frequency $\Omega$ to be equal to the mean frequency $\overline{\omega}_i$. Hence we conclude that the synchronisation frequency is determined solely by the mean of the distribution and not its shape.

\section{Synchronisation in directed ring network}

To construct a general system whose synchronisation frequency can be changed by altering the shape of the distribution, we consider a ring topology in the oscillators where each oscillator interacts with its immediate neighbour only and the last oscillators interacts with the first oscillator. This is described mathematically as
\begin{equation}
\dot{\theta_i}=\omega_i + K \sin(\theta_{i+1}-\theta_{i}) ~~~ \forall i=1,2,...,N
\label{eq: Ring Kuramoto}
\end{equation}
with the boundary condition $\theta_{N+1}=\theta_N$.

As in the previous case, the natural frequencies $\omega_i$'s are chosen from a general distribution $g(\omega)$. From here on, we assume without any loss of generality that the mean of $g(\omega)$ is zero. This is because in case, the mean is different, we can shift to a frame rotating with the mean frequency without changing the dynamics of the system. In such a rotating frame, the apparent frequencies would seem to have a mean zero.

\subsection{Numerical Results on synchronisation condition}

To analyse synchronisation phenomenon in the directed ring network, numerical simulations are performed by RK4 method for numerical integration at double precision. Ensembles of oscillators with random natural frequencies are constructed \cite{neal2003slice,robert2004monte}. Natural frequencies are chosen from symmetric (Gaussian and uniform) and asymmetric ($\chi^2$ and log-normal) distributions. For the simulations the distributions are chosen to have mean zero and sampling is performed using inverse Fourier and rejection methods \cite{ross2006simulation,aitchison1957lognormal}.

For each simulation, the time evolution of the order parameter \cite{kuramoto2003chemical}
\begin{equation}
\mathcal{O} = r e^{\iota \psi} = \frac{1}{N} \sum_{j=1}^{N} e^{\iota \theta_j}
\label{eq:Order Parameter}
\end{equation}
is also computed. Here the coherence parameter
\begin{equation}
r(t)=\frac{1}{N}\sqrt{(\sum_{i=1}^{N} \cos \theta_i)^2+(\sum_{i=1}^{N} \sin \theta_i)^2}.
\end{equation}
lies between 0 and 1 and measures the phase coherence. If all the oscillators move in a tight clump, the phases are almost the same and $r \approx 1$; whereas if all the oscillators are scattered, then $r \approx 0$.


The results obtained for symmetric $g(\omega)$ are similar to those of the mean field Kuramoto oscillators. For very low $K$ values we obtain $r(t) \rightarrow 0$ as $t \rightarrow \infty$.  For this case the value of $\dot{\theta}$ averaged over all oscillators fluctuates with time; and we see no synchronisation.For a large value of $K$, $r(t) \rightarrow 1$ as $t \rightarrow \infty$ and the system gets synchronised to the mean of the natural frequency distribution, which is zero in this case.

\begin{figure}[htbp]
\begin{subfigure}{0.5\textwidth}
\includegraphics[width=0.8\textwidth]{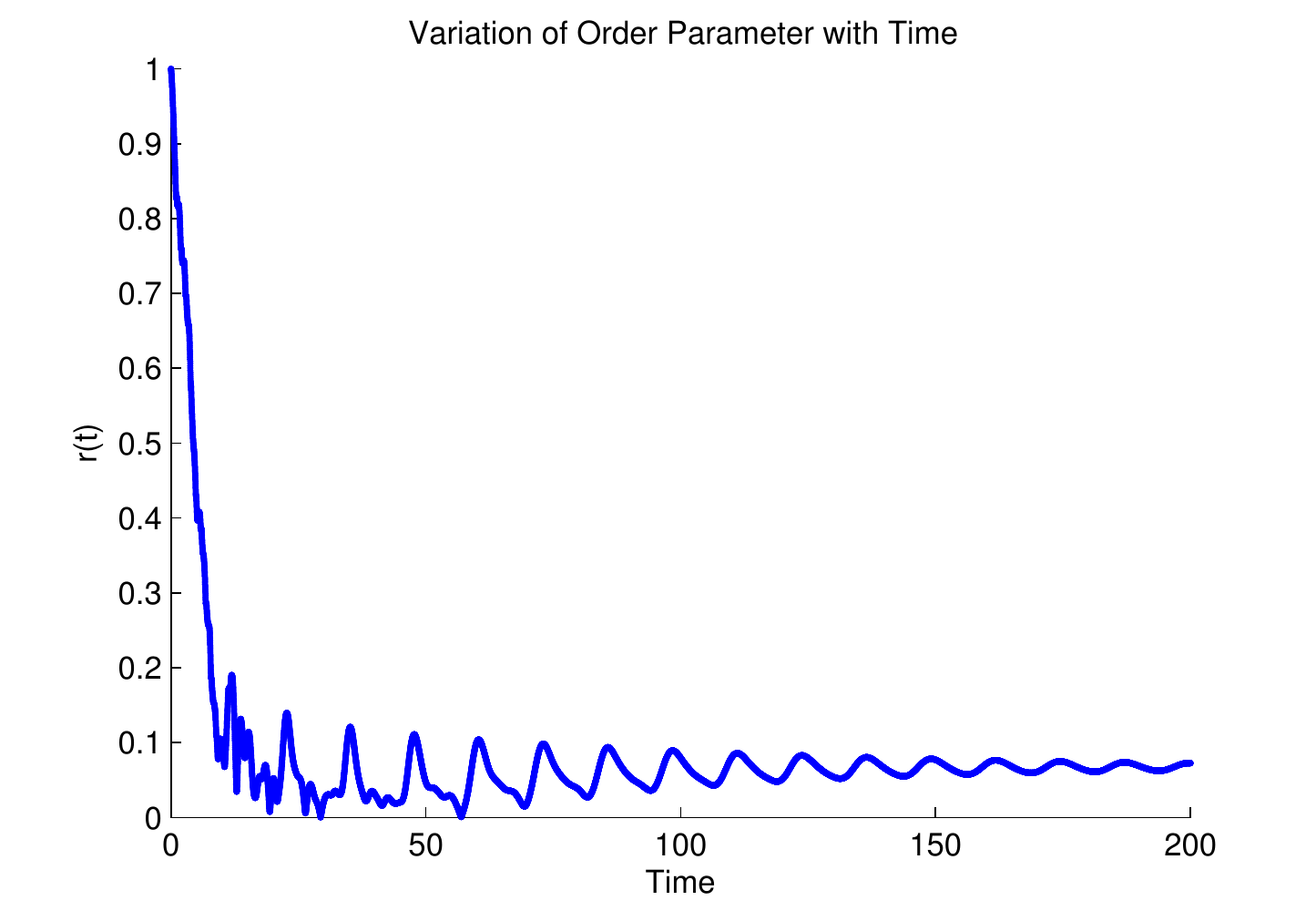}
\caption{}
\label{fig: Order Parameter}
\end{subfigure}
\begin{subfigure}{0.5\textwidth}
\includegraphics[width=0.8\textwidth]{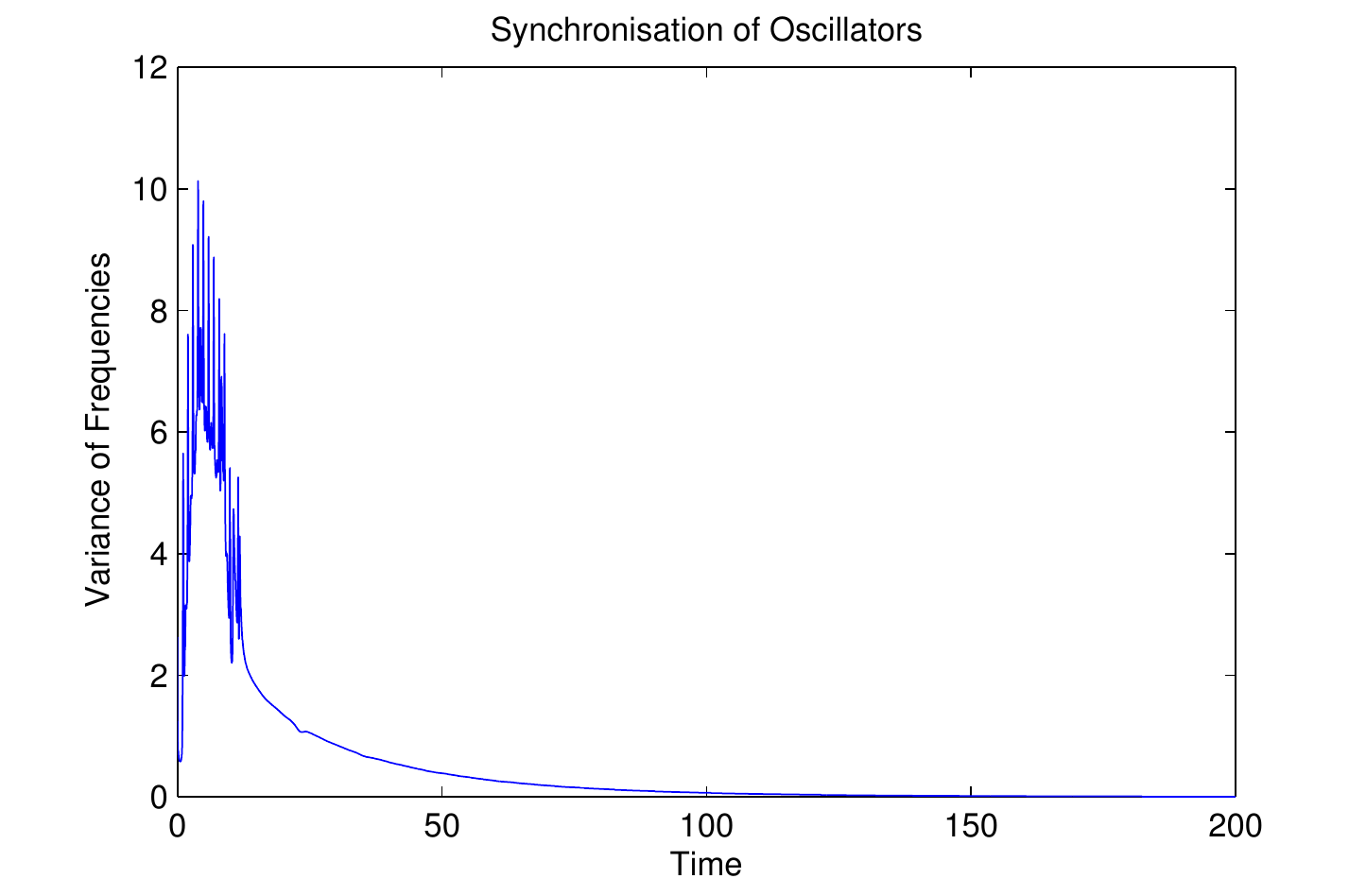}
\caption{}
\label{fig: Variance convergence}
\end{subfigure}
\begin{subfigure}{0.5\textwidth}
\includegraphics[width=0.8\textwidth]{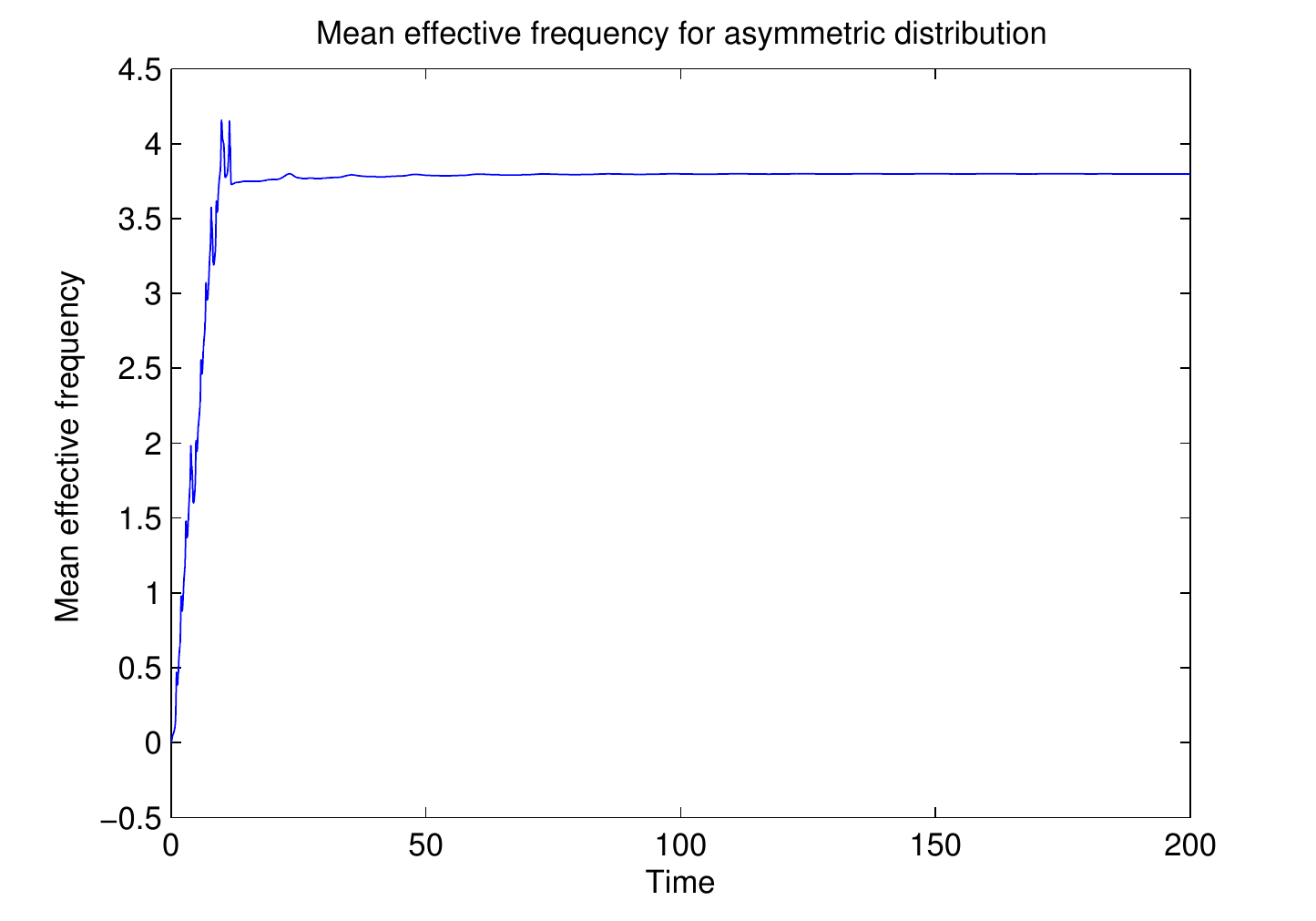}
\caption{}
\label{fig: Mean Frequency convergence}
\end{subfigure}
\begin{subfigure}{0.5\textwidth}
\includegraphics[width=0.8\textwidth]{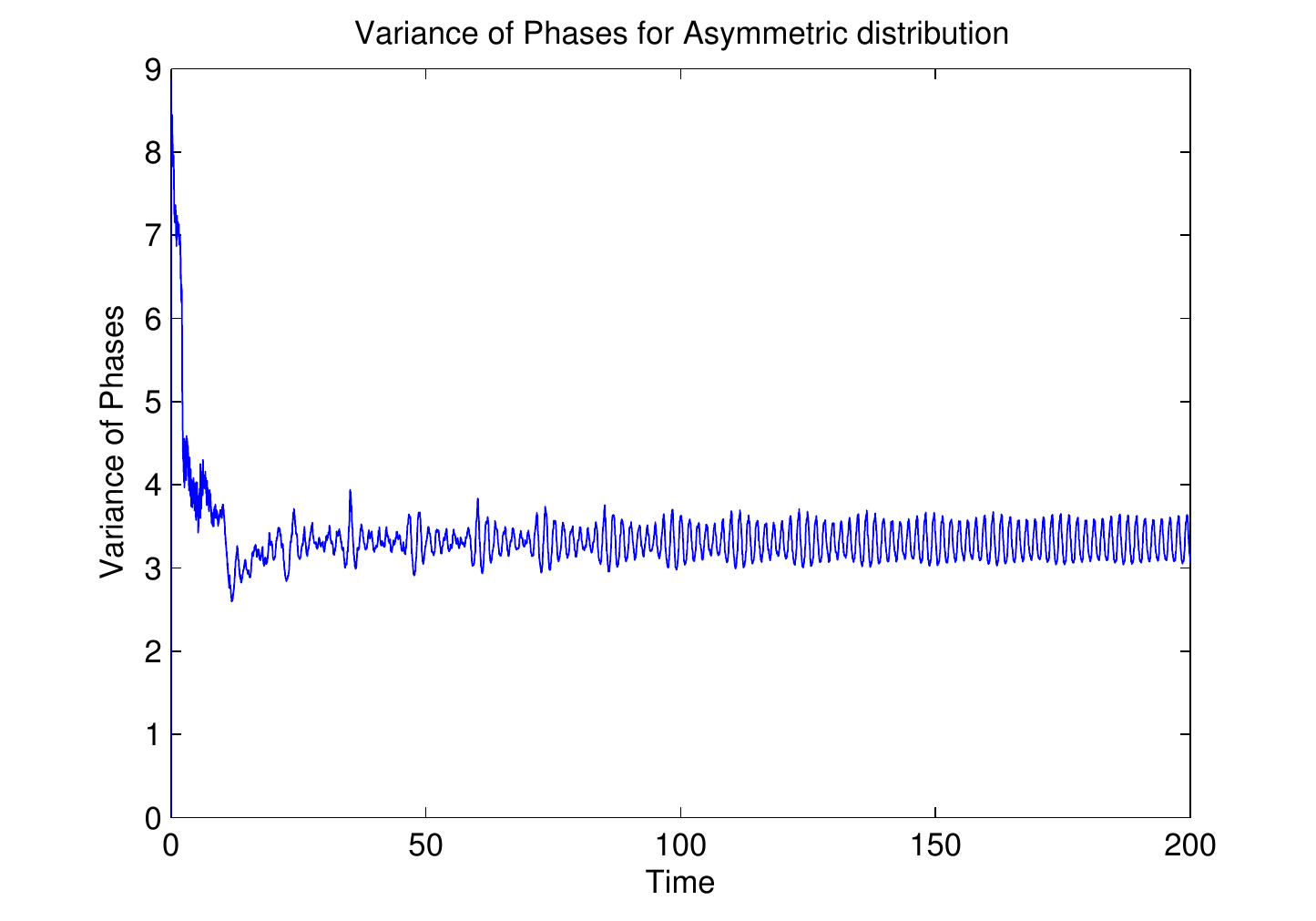}
\caption{}
\label{fig: Phase Variance}
\end{subfigure}
\caption{Numerical results of simulations with asymmetric distribution of natural frequencies. The network gets synchronised as seen by the variance of frequency decaying down to zero (in Figure(b)). For the same simulation, the coherence parameter (in a) attains avery low value and the mean effective frequency (in Figure(c)) --- which is the common synchronisation frequency for asymptotically low variance --- settles to a value away from the mean of the natural frequencies. Figure(d) shows the variance of the phases which stabilises to a finite value showing spread of oscillators in the synchronised state.}
\end{figure}

For very low and very high values of $K$, the behaviour of the asymmetric system is much like the that of the symmetric one. However for intermediate values of $K$, the asymmetric oscillators become phase locked even for very low value asymptotic value of $r(t)$ (values less than 0.1) (Figure \ref{fig: Order Parameter}). This can be clearly seen from the fact that variance of $\dot{\theta_i}$ tends to zero with time (Figure \ref{fig: Variance convergence}) and that the difference between the phases of the oscillators becomes approximately constant. What is more interesting is that in this state the effective synchronisation frequency of the oscillators shifts from the mean of their natural frequencies (Figure \ref{fig: Mean Frequency convergence}).

The phenomenon of synchronisation at low values of $r$ for asymmetric distribution points towards phases being spread out as the oscillators get synchronised. While the possibility of such `spread-out synchronised states' cannot be denied in symmetric distributions, our simulations suggest that such a state is seen only in networks with asymmetric natural frequency distribution.

\subsection{Theoretical analysis of synchronisation condition}

To obtain the condition for synchronisation, we assume that each oscillator moves with an effective frequency $\dot{\theta_i}=\Omega$. We then invert the equation \eqref{eq: Ring Kuramoto} to obtain
\begin{equation}
\sin^{-1}\left(\frac{\Omega-\omega_i}{K}\right)=\theta_{i+1}-\theta_{i} ~~~ \forall i=1,2,...,N
\end{equation}
Summing over all oscillators cancels out the right hand side completely to yield the condition for synchronisation to be
\begin{equation}
\sum_{i=1}^{N}\sin^{-1}\left(\frac{\Omega-\omega_i}{K}\right)=0.
\label{eq: Synchronisation Condition}
\end{equation}

Now to distinguish between the synchronisation frequencies of symmetric and asymmetric natural frequency distributions, we use the Taylor expansion of $\sin^{-1}x$ around $x=0$
\begin{equation}
\sin^{-1}x=\sum_{n=0}^{\infty} \frac{\Gamma\left(n+\frac{1}{2}\right)}{\sqrt{\pi}\left(2n+1\right)n!} x^{2n+1}.
\label{eq: Sine Inverse Expansion}
\end{equation}
Note that this expansion is valid for the small phase differences $\theta_{i+1}-\theta_i$ and hence for large enough $K$ values.

Denoting the coefficients $x^{2n+1}$ to be $c_{2n+1}$, and taking $x=\left(\frac{\Omega-\omega_i}{K}\right)$, we have
\begin{equation}
\sin^{-1}\left(\frac{\Omega-\omega_i}{K} \right) = \sum_{n=1}^{\infty} c_{2n+1} \left(\frac{\Omega-\omega_i}{K}\right)^{2n+1}
\end{equation}
or explicitly
\begin{equation}
\sin^{-1}\left(\frac{\Omega-\omega_i}{K} \right) = c_1\left(\frac{\Omega-\omega_i}{K} \right) + c_3{\left(\frac{\Omega-\omega_i}{K} \right)}^3 + c_5{\left(\frac{\Omega-\omega_i}{K} \right)}^5 \ldots \nonumber
\end{equation}

Applying binomial expansion to each term in the RHS, we have
\begin{equation}
\sin^{-1}\left(\frac{\Omega-\omega_i}{K} \right) = \sum_{n=1}^{\infty} c_{2n+1} \sum_{j=0}^{2n+1}(-1)^j~^{2n+1}C_{j} \left(\frac{\Omega}{K}\right)^{2n+1-j} \left(\frac{\omega_i}{K}\right)^{j} \nonumber
\end{equation}

Summing over all oscillators and using equation \eqref{eq: Synchronisation Condition}, we obtain the condition for synchronised state to be
\begin{equation}
\sum_{n=1}^{\infty} c_{2n+1} \sum_{j=0}^{2n+1}(-1)^j~^{2n+1}C_{j} \left(\frac{\Omega}{K}\right)^{2n+1-j} \sum_{i=1}^{N}\left(\frac{\omega_i}{K}\right)^{j} = 0.
\end{equation}

Collecting the terms of sums of same powers of $\omega_i$'s together, the equation can be written as an infinite series in the moments of the natural frequency distribution of the oscillators equated to zero
\begin{eqnarray}
0 &=& \left( c_1 \left(\frac{\Omega}{K}\right) + c_3 \left(\frac{\Omega}{K}\right)^3 + c_5 \left(\frac{\Omega}{K}\right)^5 + \ldots \right) \mu_0 \nonumber \\  &-& 
\left( c_1 + {^3C_1} c_3 \left(\frac{\Omega}{K}\right)^2 + {^5C_1} c_5 \left(\frac{\Omega}{K}\right)^4 + \ldots \right) \mu_1 \nonumber \\ &+& 
\left( {^3C_2} c_3 \left(\frac{\Omega}{K}\right) + {^5C_2} c_5 \left(\frac{\Omega}{K}\right)^3 + \ldots \right) \mu_2 \nonumber \\ &-& 
\left( c_3 + {^5C_3} c_5 \left(\frac{\Omega}{K}\right)^2 + \ldots \right) \mu_3 \nonumber \\ &+& 
\ldots
\label{eq: Taylor Expansion}
\end{eqnarray}
where
\begin{equation}
\mu_j=\sum_{i=1}^{N} \left(\frac{\omega_i}{K}\right)^j.
\end{equation}

From equation \eqref{eq: Taylor Expansion}, we require that each of the term of the series must drop out to zero for synchronisation. Now, for a symmetric distributions, $\mu_j=0$ for all odd $j$'s. We also note that the coefficients of $\mu_j$ for all the remaining terms are of the form contain odd powers $\frac{\Omega}{K}$ only. Hence we can see that $\Omega=0$ is a stable solution to the equation for a symmetric $g(\omega)$ with mean zero.

For an asymmetric distribution of natural frequencies however, at least one of the odd moments of the distribution is non-zero. That being the case, $\Omega=0$ is no longer a general solution as there are terms independent of $\left(\frac{\Omega}{K}\right)$ in the coefficients of $\mu_j$ for odd $j$. Hence the terms corresponding to those non-zero odd moments do not vanish on substituting $\Omega=0$. Therefore for asymmetric distributions the synchronisation frequency differs from the mean of the frequency distribution $g(\omega)$.

An estimate of the deviation of synchronisation frequency for slightly asymmetric distributions can obtained by analysing the stability of roots of the equation \eqref{eq: Synchronisation Condition}. For this we invoke a general result concerning roots of continuous and differentiable functions. Consider a continuous and differentiable function $p(x)$ which has a simple root at $x=\alpha$. If a small perturbation is introduced in the function in the form
\begin{equation}
z(x)=p(x)+\epsilon q(x)
\end{equation}
where $q(x)$ is another continuous and differentiable function and $\epsilon$ is the size of the perturbation, then for small perturbations, the root $x=\alpha^{\prime}$ of $z(x)$ corresponding to the root $x=\alpha$ of $p(x)$ is given as \cite{atkinson2008introduction}
\begin{equation}
\alpha^{\prime} = \alpha - \frac{q(\alpha)}{p^{\prime}(\alpha)}\epsilon.
\label{eq: Root Stability}
\end{equation}

For our purposes, synchronisation frequency of the oscillators is nothing but the roots of the RHS of equation \eqref{eq: Synchronisation Condition}. For a completely symmetric distribution, the function contains $\mu_j$ terms with even $j$ values only. Now, if a small asymmetry is introduced in the distribution by adding a small odd $i$-moment; it effectively implies addition of a small $\mu_i$ to the equation \eqref{eq: Synchronisation Condition}. This term can be considered a small perturbation with $\epsilon=\mu_i$ to the function with even $\mu_j$ 's whose one of the simple roots is known to be zero.

Hence considering a natural frequency distribution $g(\omega)$ which is asymmetric only due to a small odd $i$-moment, the synchronisation frequency is obtained by direct substitution in equation \eqref{eq: Root Stability} to give
\begin{equation}
\Omega = \frac{c_i \mu_i}{\sum\limits_{j=0}^{\infty}(2j+1)~c_{2j+1}~\mu_{2j}}
\label{eq: Synchronisation Frequency}
\end{equation}
where $c_j$ is the coefficient of $x^{j}$ in the expansion of $\sin^{-1} x$ (equation \eqref{eq: Sine Inverse Expansion}).

Some conclusions can be immediately drawn from equation \eqref{eq: Synchronisation Frequency}. Firstly, as the values of $c_j$ decreases with increasing $j$, the effect of asymmetry on shifting of the synchronisation frequency is maximum if asymmetry is introduced by increasing $\mu_3$ or the skewness of $g(\omega)$. Secondly as the denominator of the expression contains weighted sum of even moments of the distribution, we can conclude that a greater shift in synchronisation frequency will be observed for a sharper distribution of natural frequencies. Thirdly, in general if the coupling strength $K$ decreases, the shift in the synchronisation frequency increases. This is in accordance with the numerical simulations obtained where the synchronisation frequency of the oscillators is seen to converge to the mean as the coupling becomes very large.

\section{Conclusion}

In this article we study the effect of symmetry of natural frequency distribution on the synchronisation frequency of Kuramoto oscillators. After establishing that synchronisation frequency of a complete network of oscillators is unaffected by symmetry of the distribution of natural frequencies, we construct a ring network of oscillators to observe a shift in synchronisation frequency for asymmetric natural frequency distribution and an intermediate range of coupling strength. We analytically show that oscillators can synchronise to the mean of the frequency distribution only for symmetric distributions. As asymmetry is introduced in a synchronised symmetric system; the synchronisation frequency gradually shifts way from the mean. An estimate of the shift for small asymmetry is also given. The results also qualitative predict the increase in the shift with reduction in coupling strength which is also seen in numerical simulations.



Although we have been able to come up with a network of Kuramoto oscillators whose synchronisation frequency can be changed by introducing asymmetries in the natural frequency distribution, a few questions still remain open and unanswered. These include estimation of synchronisation frequency for a more general asymmetric distribution and characterisation of phase difference between the oscillators for the non-collapsed phase locked state. We hope that future research will throw some light on the unaddressed issues and help us characterise the effect of asymmetries in a more efficient way.

\newpage
\bibliographystyle{unsrt}
\bibliography{Kuramoto}

\begin{thebibliography}{10}

\bibitem{buck1988synchronous}
J.~Buck.
\newblock Synchronous rhythmic flashing of fireflies. ii.
\newblock {\em Quarterly Review of Biology}, pages 265--289, 1988.

\bibitem{neda2000physics}
Z.~N{\'e}da, E.~Ravasz, T.~Vicsek, Y.~Brechet, and A.L. Barab{\'a}si.
\newblock Physics of the rhythmic applause.
\newblock {\em Physical Review E}, 61(6):6987, 2000.

\bibitem{gerstner1995time}
W.~Gerstner.
\newblock Time structure of the activity in neural network models.
\newblock {\em Physical Review E}, 51(1):738, 1995.

\bibitem{PhysRevE.57.1563}
Kurt Wiesenfeld, Pere Colet, and Steven~H. Strogatz.
\newblock Frequency locking in josephson arrays: Connection with the kuramoto
  model.
\newblock {\em Phys. Rev. E}, 57:1563--1569, Feb 1998.

\bibitem{wiesenfeld1996synchronization}
Kurt Wiesenfeld, Pere Colet, and Steven~H Strogatz.
\newblock Synchronization transitions in a disordered josephson series array.
\newblock {\em Physical review letters}, 76(3):404, 1996.

\bibitem{kuramoto2003chemical}
Y.~Kuramoto.
\newblock {\em Chemical oscillations, waves, and turbulence}.
\newblock Dover Pubns, 2003.

\bibitem{wiener1966nonlinear}
N.~Wiener.
\newblock Nonlinear problems in random theory.
\newblock {\em Nonlinear Problems in Random Theory, by Norbert Wiener, pp. 142.
  ISBN 0-262-73012-X. Cambridge, Massachusetts, USA: The MIT Press, August
  1966.(Paper)}, 1, 1966.

\bibitem{wiener1948cybernetics}
N.~Wiener.
\newblock Cybernetics.
\newblock {\em Scientific American}, 179(5):14, 1948.

\bibitem{strogatz2000kuramoto}
S.H. Strogatz.
\newblock From kuramoto to crawford: exploring the onset of synchronization in
  populations of coupled oscillators.
\newblock {\em Physica D: Nonlinear Phenomena}, 143(1):1--20, 2000.

\bibitem{neal2003slice}
Radford~M Neal.
\newblock Slice sampling.
\newblock {\em Annals of statistics}, pages 705--741, 2003.

\bibitem{robert2004monte}
Christian~P Robert and George Casella.
\newblock {\em Monte Carlo statistical methods}, volume 319.
\newblock Citeseer, 2004.

\bibitem{ross2006simulation}
S.M. Ross.
\newblock {\em Simulation}.
\newblock Elsevier Academic Press, Amsterdam, 2006.

\bibitem{aitchison1957lognormal}
J.~Aitchison and JAC Brown.
\newblock The lognormal distribution. university of cambridge, department of
  applied economics, monograph no. 5, 1957.

\bibitem{atkinson2008introduction}
Kendall~E Atkinson.
\newblock {\em An introduction to numerical analysis}.
\newblock John Wiley \& Sons, 2008.

\end{thebibliography}

\end{document}